\documentclass[12pt,draftcls,onecolumn]{IEEEtran}     

\IEEEoverridecommandlockouts                              
\usepackage{amsfonts}
\usepackage{amsmath}
\usepackage{graphicx}      
\usepackage{float} 
\usepackage{psfrag} 
\usepackage{mathtools}
\usepackage{url}
\usepackage{cite}

\title{\LARGE  \bf Dual-Update Data-Driven Control of Deformable Mirrors  Using Walsh Basis Functions}

\author{Aleksandar Habe$\text{r}^1$,  and  Thomas Bifan$\text{o}^{2}$
	\thanks{$^1$ Department of Manufacturing and Mechanical Engineering Technology, College of Engineering Technology, Rochester Institute of Technology, 78 Lomb Memorial Drive, Rochester, NY 14623, USA. \\
	 $^{2}$  Mechanical Engineering Department, Boston University, 8 Saint Mary’s Street, Boston, MA 02215, USA. \\
Emails: awhmet@rit.edu, aleksandar.haber@gmail.com}}

\begin{document}
\maketitle
\thispagestyle{empty}
\pagestyle{empty}
\maketitle
\begin{abstract}
In this paper, we develop a novel data-driven method for Deformable Mirror (DM) control. The developed method updates both the DM model and DM control actions that produce desired mirror surface shapes. The novel method explicitly takes into account actuator constraints and couples a feedback control algorithm with an algorithm for recursive estimation of DM influence function models. In addition to this, we explore the possibility of using Walsh basis functions for DM control.  By expressing the desired and observed mirror surface shapes as sums of Walsh pattern matrices, we formulate the control problem in the 2D Walsh basis domain.  We thoroughly experimentally verify the developed approach on a 140-actuator MEMS DM, developed by Boston Micromachines. Our results show that the novel method produces the root-mean-square surface error in the $14-40$ nanometer range. These results can additionally be improved by tuning the control and estimation parameters. The developed approach is also applicable to other DM types, such as for example, segmented DMs. 
\end{abstract}

\section{Introduction}

Deformable Mirrors (DMs) are one of the main components of Adaptive Optics (AO) systems~\cite{tyson2010principles,Roddier1999}. A typical DM consist of a reflective optical surface that is deformed by a set of actuators. By precisely shaping the surface of a DM, we can compensate for wave-front aberrations in AO systems~\cite{manetti2014vlt,Vdovin1995,Ravensbergen2012,madec2012overview,polo2013linear,horenstein1999real,kuiper2018advances,haber2013identification,haber2016framework,haber2013predictive,saathof2015actuation,haber2020modelingHaberVerhaegen,chiuso2009dynamic,mocci2020pi}.

In this paper, we consider the problem of developing control algorithms for DMs. There are a large number of approaches for DM control. A complete survey of all the methods goes well beyond the length limits of this manuscript. Consequently, we briefly mention only the most relevant or recent approaches. Most of the DM control approaches are based on the following control paradigm~\cite{Haber:13,Fernandez2003,vogel2010modeling,vogel2011modeling,Polo:12}. First, a model of the DM (influence function matrix) is estimated before the correction process. Then, such a model is either directly inverted, or it is used together with an iterative feedback control algorithm to calculate the DM control actions. That is, in most of the feedback control approaches, only control actions (control voltages) for the DM are updated on the basis of the observed wave-front aberrations, and the mirror model is kept constant. However, this approach that is based on linear and time-invariant DM models, might not be able to produce satisfactory wave-front correction performance in at least two scenarios that are encountered in practice. The first scenario is when the DM behavior changes over time. For example, an environment in which a DM operates might induce temperature fluctuations that can create thermo-mechanical deformations of mirror components and other associated phenomena that can significantly alter the DM dynamics~\cite{gu2019thermal,blaurock2005structural,buleri2019structural,xue2013research,Kasprzack:13,haber2020modeling,haber2021modeling,haber2013predictive,saathof2015actuation,habets2016multi,haber2013identification}. This is especially the case for optical systems operating in space and for optical systems operating with high-power laser sources, where the absorbed heat increases the temperature of optical components. The second scenario is when the DM control actions significantly deviate from an initial linearization point (determined for example by bias control actions) that is used to obtain the linear model. Namely, most of the developed DM devices and prototypes exhibit a nonlinear behavior for sufficiently large magnitudes of control actions that are necessary to produce wave-front shapes with larger Peak-to-Valleys (PVs). That is, to correct for certain wave-front shapes, control actions need to significantly deviate from the nominal control actions that are used to obtain linearized DM models. In both of these scenarios, the performance of the DM control algorithms developed on the basis of time-invariant and linear models estimated for nominal operating conditions will gradually degrade.

On the other hand, it is a very well-known fact in control theory literature that control methods that dynamically update systems' models on the basis of available observations are able to cope with time-varying dynamics, model uncertainties, and system nonlinearities. Furthermore, even in the case of linear systems, adaptive control methods are more accurate, faster, and in many cases, more optimal than classical feedback algorithms. All these facts motivate the development of a \textit{dual-update} control algorithm in this paper. Under the term ``dual-update'' control algorithm, we understand a control algorithm that is able to update both control actions and a model (influence function) during DM operation. Here is it should be emphasized that these types of algorithms are often referred to as adaptive algorithms in control theory literature. However, since the word ``adaptive'' has different meanings in optics and control theory communities, we use the word ``dual-update'', not to confuse an interested reader.

To the best of our knowledge, adaptive control approaches for DM control have received less attention and interest compared to the classical optimal control approaches~\cite{kulcsar2012minimum,mocci2020pi}. Thus, the approach proposed in~\cite{zou2009high} iteratively calibrates a DM influence function during the correction process. The approach developed in~\cite{huang2015high} uses a recursive least-squares method to dynamically estimate DM influence function during DM operation. The main limitation of the approaches developed in~\cite{huang2015high,zou2009high} is that they do not explicitly take into account actuator saturation. A reliable control method has to take into account actuator physical constraints and has to be able to deal with actuator saturation. Recently, in~\cite{haber2021general}, we have developed a DM control approach that updates the DM model on the basis of batches of observed wave-front data. This method takes into account actuator saturation. However, once the DM model is updated, the control is performed in an open-loop setting. Consequently, this method is not suitable for online control, where adaptive DM model updating is coupled with feedback control. Apart from these methods, adaptive filters have been used in~\cite{tesch2013receding} to predict and control wave-front disturbances. 

A widely used approach for DM control is to express the observed wave-front using a Zernike modal basis. In this way, we implicitly perform the spatial model-order reduction of the DM control problem (the control problem is transformed from the spatial domain to the modal basis domain). On the other hand, for certain mirror types and for desired wave-front shapes consisting of steep peaks and valleys, Walsh basis functions might be a more suitable option than the Zernike basis functions~\cite{wang2011utility,wang2009wavefront,wang2012high}. Apart from this, in the general case, since Walsh basis functions are orthogonal, they can be used instead of Zernike polynomials in classical adaptive optics applications. 

In this paper, we develop a novel dual-update DM control approach. The developed approach explicitly takes into account actuator constraints and it couples a feedback control algorithm with a recursive estimation of the DM influence function model. In this way, we are able to dynamically update the DM model and at the same time compute control actions that produce the desired shape. Furthermore, by expressing the desired and observed mirror surface shapes as sums of Walsh pattern matrices, we formulate the control problem in the 2D Walsh basis domain. We experimentally verify the developed approach on a 140 actuator MEMS DM, developed by Boston Micromachines.

The main contributions of this paper are summarized in the sequel. We propose a novel dual-update control approach for accurate DM control. Furthermore, in contrast to other dual-update or adaptive DM control approaches proposed in the literature, our approach explicitly takes into account actuator saturation. In this way, we can avoid the loss of performance that comes from actuator saturation. In addition, in this paper, we thoroughly experimentally investigate the performance of Walsh basis functions for DM control. This is important since only a handful of articles have explored the possibility of using Walsh basis functions for wave-front reconstruction and control~\cite{wang2011utility,wang2009wavefront,wang2012high}, and the true potential of using Walsh basis functions for DM control is largely unexplored. It should be emphasized here that although we have performed experiments on a continuous face sheet DM, the developed control approach is applicable to other DM types. The developed approach is especially suitable for segmented DMs. Finally, we can easily modify the develop control algorithm to use Zernike basis functions instead of the Walsh basis functions.

This paper is organized as follows. In Section~\ref{WalshSection}, we present the procedure for approximating the mirror surface shape as a sum of Walsh pattern matrices. In Section~\ref{ControlMethodSection}, we present the control method. In Section~\ref{experimentalResultsSection}, we present the experimental results. In Section~\ref{SectionConclusion}, we present conclusions and briefly discuss future work.

\section{Mirror deformation representation using Walsh pattern matrices}
\label{WalshSection}

In this section, we present a simple numerical procedure for approximating the mirror surface shape as a sum of Walsh pattern matrices.

Mirror surface deformation is usually represented by a matrix. That is, every entry of this matrix is a mirror deformation at a fixed spatial location. We refer to this matrix as the \textit{mirror deformation matrix}. Let $W\in \mathbb{R}^{n\times n}$ be the mirror deformation matrix, where $n$ is the total size (measured in pixels) along $x$ and $y$ dimensions of the observed mirror surface. We decompose this matrix as follows 
\begin{align}
W \approx \sum_{p=1}^{M}\sum_{q=1}^{M} a_{p,q} Z_{p,q},
\label{matrixDecomposition}
\end{align}
where $a_{p,q}\in \mathbb{R}$ are coefficients, and $Z_{p,q} \in \mathbb{R}^{n\times n}$ are Walsh pattern matrices with the entries that can either be $-1$ or $1$. The number $n$ should be selected such that $n=2^{V}$, where $V$ is a user-selected positive integer. The total number of the Walsh pattern matrices in \eqref{matrixDecomposition} is equal to $M^{2}$, where $M \le n$. We form the Walsh pattern matrices using Walsh basis functions. Here it should be emphasized that since this paper presents a proof of concept and due to mathematical simplicity and brevity, we use Walsh basis functions defined over a square domain. The mirror surface shape can also be represented using Walsh basis functions defined over a circular domain (polar Walsh basis functions)~\cite{hazra1986far}. Consequently, the developed method can easily be used in the case of circular correction domains. However, even without using polar Walsh basis functions, with some modifications, the used approach that is based on square-domain Walsh basis functions, can be used for wave-front correction over circular correction domains. On the other hand, the control approach that is developed in Section~\ref{ControlMethodSection} is practically independent of the type of basis functions for expanding the mirror surface shape. Thus, instead of using Walsh basis functions, we can also use Zernike basis functions in the developed control approach.  

In the sequel, we first introduce a procedure for constructing the Walsh pattern matrices $Z_{p,q}$. Then, we introduce a procedure for computing the coefficients $a_{p,q}$. First, we choose the constant $V$. We select the constant $V$ such that a deformation submatrix with the dimensions of $2^{V}$ by $2^{V}$ pixels is within the limits of the maximal active mirror surface area that is observable by the used sensor (more details about the sensor used in our experiments can be found in Section~\ref{experimentalResultsSection}). In our case, we use $V=8$ or $V=9$, giving us the submatrices with the dimensions of $256$ by $256$, and $512$ by $512$, respectively (see Section~\ref{experimentalResultsSection} for more details).

 Let the entries of the vector $\boldsymbol{\gamma}_{i}^{(V)}\in \mathbb{R}^{n}$, $n=2^{V}$, represent the values of the Walsh function $i$ of the order $V$. Here the index $i$ takes the values from $1$ to $2^{V}$. For example for $V=2$, the vectors $\boldsymbol{\gamma}_{i}^{(2)}$, $i=1,2,3,4$, representing the Walsh basis functions take the following form 
\begin{align}
\boldsymbol{\gamma}_{1}^{(2)}=\begin{bmatrix}1 \\ 1 \\ 1 \\ 1 \end{bmatrix}, \;\; \boldsymbol{\gamma}_{2}^{(2)}=\begin{bmatrix}1 \\ 1 \\ -1 \\ -1 \end{bmatrix}, \boldsymbol{\gamma}_{3}^{(2)}=\begin{bmatrix}1 \\ -1 \\ -1 \\ 1 \end{bmatrix}, \;\; \boldsymbol{\gamma}_{4}^{(2)}=\begin{bmatrix}1 \\ -1 \\ 1 \\ -1 \end{bmatrix}.
\label{WalshFunctions}
\end{align}
The vectors  $\boldsymbol{\gamma}_{i}^{(V)}$ can easily be constructed by using the MATLAB function $\text{hadamard}(\cdot)$. The rows of a matrix returned by this function represent Walsh basis functions. However, the Walsh basis functions that are represented by the rows of this matrix are not arranged in increasing order. Consequently, the rows of the matrix returned by the function $\text{hadamard}(\cdot)$ need to be permuted. For more details, see the MATLAB tutorial page~\cite{matlabTutorial1} explaining the construction process of the Walsh basis functions. 

Once the vectors $\boldsymbol{\gamma}_{i}^{(V)}$ are constructed, for selected $V$, we use the following procedure to construct the Walsh pattern matrices. First, we need to select the constant $M$. While selecting the constant $M$ we have to keep in mind several competing factors. Generally speaking, we have to make a trade-off between representation accuracy that is increased by increasing $M$, and dimensions of the matrices of the control algorithm that increase with the factor of $M^2$. Larger matrix dimensions increase the computational and memory complexities of the decomposition process, as well as the computational and memory complexities of the control and estimation algorithms that are introduced in Section~\ref{ControlMethodSection}. We have tested the control algorithm for $M$ up to $120$ and this value is sufficient for desired shapes used in our experiments. Our lab computer has $64$ GB RAM, and for $M \ge 120$, MATLAB programming language that is used to control the DM, runs out of memory. Larger values of $M$ are possible if the control algorithm is implemented in a more memory-efficient way. One of the possible pathways to decrease the computational and memory complexities is to exploit the structure of the control matrices using approaches similar to the approaches developed in~\cite{massioni2011fast,massioni2015approximation,haber2016framework,Haber:13mhe,haber2018sparsity,haber2014sparseLyapunov,haber2014subspace,sinquin2018tensor,monchen2018recursive,cerqueira2021sparse}. However, this requires additional research, implementation, and testing efforts that are left for future research.

\begin{figure}[H]
\centering 
\includegraphics[scale=0.25,trim=0mm 0mm 0mm 0mm ,clip=true]{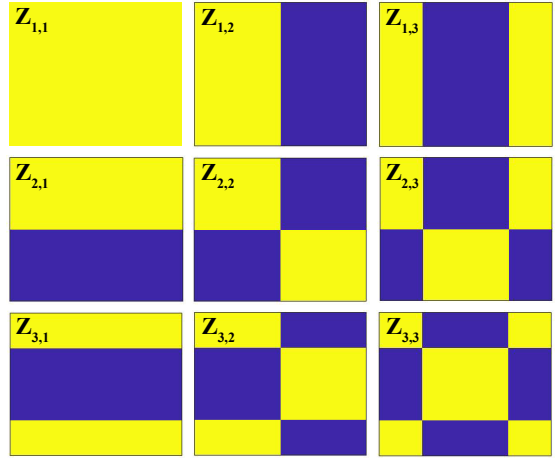}
\caption{2D Walsh pattern matrices. The yellow and blue colors correspond to $1$ and $-1$ values, respectively.}
\label{fig:Graph1}
\end{figure}

Once we have selected $V$ and $M$, we can proceed with the construction of the pattern matrices. We perform the following steps:

\begin{enumerate}

\item \textbf{Step 1:} In this step, we construct the matrices $Z_{1,q}\in \mathbb{R}^{n}$, where $q=1,2,\ldots, M$, and $n=2^{V}$. The matrix $Z_{1,q}$ is constructed by transposing the vector $\boldsymbol{\gamma}_{q}^{(V)}$ and by stacking the newly formed row vectors on top of each other $n$ times:
\begin{align}
Z_{1,q}=\begin{bmatrix}  \Big(\boldsymbol{\gamma}_{q}^{(V)}\Big)^{T}  \\ \Big(\boldsymbol{\gamma}_{q}^{(V)}\Big)^{T} \\ \vdots  \\ \Big(\boldsymbol{\gamma}_{q}^{(V)}\Big)^{T}   \end{bmatrix}.
\label{matrixZ1}
\end{align}

\item \textbf{Step 2:} In this step, we construct the matrices $Z_{q,1}$, where $q=1,2,\ldots, M$. The matrices $Z_{q,1}$ are  constructed by transposing the matrices $Z_{1,q}$ that are formed in step 1, that is, $Z_{q,1}=Z_{1,q}^{T}$.

\item \textbf{Step 3:} In this step, we construct the matrices $Z_{p,q}$, for the indices $p\ge 2$ and $q\ge 2$. The matrix $Z_{p,q}$ is calculated as follows 
\begin{align}
Z_{p,q}= Z_{p,1}\odot Z_{1,q}, 
\label{matrixZpq}
\end{align}
where $\odot $ is the matrix element-wise product (Hadamard matrix product).
\end{enumerate}
A few comments about this construction procedure are in order. The matrix $Z_{p,q}$ can be seen as a $p,q$ block of a large block matrix. The first block row of this matrix is constructed in Step 1. In  Step 2, we construct the first block column, where every block is a transpose of the corresponding matrix in the first block row. Then, in step 3, we construct the remaining block matrices by simply multiplying the matrices belonging to the first block column with the matrices belonging to the first block row. The constructed 2D Walsh pattern matrices are shown in Fig.~\ref{fig:Graph1} for $M=3$.

Next, we explain the decomposition process of the matrix $W$, that is, we explain how to compute the coefficients $a_{p,q}$ in \eqref{matrixDecomposition} for known $W$. First, we vectorize the equation \eqref{matrixDecomposition}. The vectorization process is done by introducing the vectorization operator $\text{vec}(\cdot)$~\cite{verhaegen2007filtering}. Let $X\in \mathbb{R}^{n\times n}$ be an arbitrary matrix with the column vectors denoted by $\mathbf{x}_{1},\mathbf{x}_{2},\ldots, \mathbf{x}_{n},\in \mathbb{R}^{n}$.  The vectorization operator is defined by 
\begin{align}
\mathbf{x}=\text{vec}(X)\coloneqq \begin{bmatrix}\mathbf{x}_{1}^{T} &\mathbf{x}_{2}^{T} &\ldots &  \mathbf{x}_{n}^{T}   \end{bmatrix}^{T}.
\label{vectorizationOperatorDefinition}
\end{align}
That is, the vectorization operator produces the vector $\mathbf{x}\in \mathbb{R}^{n^2}$ obtained by stacking column vectors $\mathbf{x}_{i} \in \mathbb{R}^{n}$ of the matrix $X$ on top of each other. By applying the vectorization operator to \eqref{matrixDecomposition}, we obtain
\begin{align}
&\text{vec}\big( W \big) \approx \sum_{p=1}^{M}\sum_{q=1}^{M} a_{p,q} \text{vec}\big( Z_{p,q} \big), 
\label{vectorizationOperator1} \\
& \mathbf{w}=\sum_{p=1}^{M}\sum_{q=1}^{M} a_{p,q} \mathbf{z}_{p,q}, \label{vectorizationOperator2}
\end{align}
where $\mathbf{w} = \text{vec}\big( W \big)$, $\mathbf{w}\in \mathbb{R}^{n^{2}}$, and $ \mathbf{z}_{p,q} =\text{vec}\big( Z_{p,q} \big)$, $ \mathbf{z}_{p,q} \in \mathbb{R}^{n^{2}}$. On the other hand, due to the fact that the pattern matrices $Z_{p,q}$ are formed on the basis of the Walsh functions, we have that 
\begin{align}
\mathbf{z}_{l,s}^{T} \mathbf{z}_{p,q} = 
    \begin{cases}
      n^2 & , \;\; l=p \land s=q \\
      0 & , \;\; l\neq p \lor  s\neq q \\
     \end{cases}     
\label{orthogonality1}
\end{align}
The property \eqref{orthogonality1} comes from the fact that the Walsh basis functions form a complete orthogonal set of functions. This property can be used to retrieve the coefficients $a_{p,q}$. Namely, from \eqref{vectorizationOperator2} and \eqref{orthogonality1}, we have 
\begin{align}
\frac{1}{n^2}\mathbf{z}_{p,q}^{T}\mathbf{w} = a_{p,q}.
\label{consequence1}
\end{align}

The equation \eqref{consequence1} enables us to construct a projection matrix that produces the coefficients of the expansion \eqref{matrixDecomposition}. Let the matrices $\Psi_{i}\in \mathbb{R}^{M\times n^{2}}$ and $\Pi\in \mathbb{R}^{M^{2}\times n^{2}}$, and the vectors $\mathbf{a}_{i}\in \mathbb{R}^{M}$ and $\mathbf{a}\in \mathbb{R}^{M^{2}}$, be defined by 
\begin{align}
\Pi=\frac{1}{n^2} \begin{bmatrix}\Psi_{1}\\ \Psi_{2} \\ \vdots \\ \Psi_{M}  \end{bmatrix},\;\; \Psi_{i} =\begin{bmatrix} \mathbf{z}_{1,i}^{T} \\  \mathbf{z}_{2,i}^{T} \\ \vdots \\  \mathbf{z}_{M,i}^{T}  \end{bmatrix}, \;\; \mathbf{a}=\begin{bmatrix}\mathbf{a}_{1}\\ \mathbf{a}_{2} \\ \vdots \\ \mathbf{a}_{M} \end{bmatrix},\;\; \mathbf{a}_{i}=\begin{bmatrix}a_{1,i} \\ a_{2,i} \\ \vdots \\ a_{M,i}  \end{bmatrix}.
\label{matrixPiVectorA}
\end{align}

The vector $\mathbf{a}$ groups all the coefficients of the expansion \eqref{matrixDecomposition}. Then, using this construction, from \eqref{consequence1} we have
\begin{align}
\mathbf{a} = \Pi \mathbf{w}.
\label{coefficientsGrouping}
\end{align}

The equation \eqref{coefficientsGrouping} shows that to express the deformation matrix $W$ as a sum of the Walsh pattern matrices, we just need to vectorize this matrix and to multiply the result with the projection matrix $\Pi$. That is, to compute the expansion coefficients $a_{p,q}$, we need to perform a single vector-matrix operation. 

\section{Control method}
\label{ControlMethodSection}

In this section, we present the control method. The basic idea of the control method is to update both the DM control actions and the DM influence function model. To develop the control method, we use the Walsh basis function expansion presented in Section~\ref{WalshSection}.

Let $W_{D}\in \mathbb{R}^{n\times n}$ be a desired mirror shape that we want to produce, and let $\mathbf{w}_{D}=\text{vec}\big(W_{D}\big)$, $\mathbf{w}_{D}\in \mathbb{R}^{n^2}$. Using \eqref{coefficientsGrouping}, we compute the desired set of coefficients as follows
\begin{align}
\mathbf{a}_{D}=\Pi \mathbf{w}_{D},
\label{desiredSetOfCoefficients}
\end{align}
where $\Pi$ is the projection matrix introduced in \eqref{matrixPiVectorA} and $\mathbf{a}_{D}\in \mathbb{R}^{M^2}$ is the vector grouping the desired coefficients.

We send control actions to the DM or observe its deformation response at discrete-time instants, denoted by $k\in \mathbb{N}_{0}$. Let $\mathbf{w}_{k} \in \mathbb{R}^{n^2}$ be the observed mirror deformation in the vectorized form, that is $\mathbf{w}_{k}=\text{vec}\big(W_{k} \big)$, where $W_{k}\in \mathbb{R}^{n\times n}$ is the observed mirror deformation matrix at the discrete-time instant $k$.  By using the projection \eqref{coefficientsGrouping}, we have 
\begin{align}
\mathbf{a}_{k}=\Pi \mathbf{w}_{k},
\label{projection22}
\end{align}
where $\mathbf{a}_{k}\in \mathbb{R}^{M^2}$ are the coefficients. We postulate the following DM model
\begin{align}
\mathbf{a}_{k+1} = Q_{k} \mathbf{g}_{k}+\mathbf{d}_{k+1},
\label{mirrorModel}
\end{align}
where $Q_{k}\in \mathbb{R}^{M^2\times r}$ is the influence matrix, and the vector $ \mathbf{g}_{k} \in \mathbb{R}^{r}$ is defined by 
\begin{align}
\mathbf{g}_{k}=\begin{bmatrix} u_{1,k}^{\beta} & u_{2,k}^{\beta} & \ldots & u_{r,k}^{\beta} \end{bmatrix}^{T},
\label{Gvector}
\end{align}
and where $u_{i,k}$, $i=1,2,\ldots, r$, is the control input applied to the $i$th actuator at the time instant $k$, $r$ is the number of DM actuators (in our case $r=140$), and $\beta=1.7420$ is an estimate of the constant of the exponential dependence between DM control actions and the observed deformation response. This estimate is obtained using the linearized least-squares method explained in~\cite{haber2021general}. The entries of the vector $\mathbf{u}_{k}\in \mathbb{R}^{r}$ are the control inputs $u_{i,k}$. The values of the control input $u_{i,k}$ are in the interval $[0,1]$, with zero corresponding to no control action, and $1$ corresponding to the maximal control action applied to the actuator $i$ (maximal control voltage applied to the actuator). The vector $\mathbf{d}_{k+1}\in \mathbb{R}^{M^2}$, that is not known \textit{a priori},  takes into account the measurement noise and unmodeled mirror behavior that is not captured by the model $Q_{k} \mathbf{g}_{k}$.

The control method consists of two steps. In the first step, we estimate an initial value of the influence matrix and compute the initial values of control inputs. These values are used to initialize the adaptive control algorithm. The adaptive control algorithm combines recursive estimation of the influence matrix with a novel control algorithm. In the sequel, we explain the first step.

\subsection{Initial estimation of the influence matrix and control actions}
\label{initialEstimation}

For a positive integer $S \ge r $ ($S$ should be larger than the number of DM actuators), we introduce the following notation for batch data matrices:
\begin{align}
A_{1:S}=\begin{bmatrix}\mathbf{a}_{1} & \mathbf{a}_{2} & \ldots & \mathbf{a}_{S}     \end{bmatrix},\;
G_{0:S-1}=\begin{bmatrix}\mathbf{g}_{0} & \mathbf{g}_{1} & \ldots & \mathbf{g}_{S-1}     \end{bmatrix},
\label{matricesDefinitionBatch}
\end{align}
where $A_{1:S}\in \mathbb{R}^{M^2 \times S}$ and $G_{0:S-1}\in \mathbb{R}^{r \times S}$. To compute the initial values, we will assume that the influence matrix is constant. Under this assumption, from \eqref{mirrorModel}, we obtain
\begin{align}
\mathbf{a}_{k+1} = Q^{(0)} \mathbf{g}_{k} + \mathbf{d}_{k+1},
\label{initialModel}
\end{align}
where $Q^{(0)} \in \mathbb{R}^{M^2\times r}$ is the initial value of the influence matrix $Q_{k}$. We use the approach presented in~\cite{haber2021general} to estimate $Q^{(0)}$. First, we generate random input signals $\mathbf{u}_{k}$, for $k=0,1,\ldots, S-1$. The entries of the input vector $\mathbf{u}_{k}$ are generated from a normal distribution with the mean of $0.5$ and variance of $0.15$. If any of the generated entries is larger (smaller) than the upper bound $1$ (lower bound $0$), then the value of this entry is set to $1$ ($0$). By forming the equations \eqref{initialModel} for $k=0,1,2,\ldots, S-1$, we obtain a batch matrix equation. From this equation, we can estimate $Q^{(0)}$ by solving the following multi-variable least-squares problem 
\begin{align}
\min_{Q^{(0)}} \left\|A_{1:S} - Q^{(0)}G_{0:S-1} \right\|_{F}^{2},
\label{solutionInitialMatrix}
\end{align}
where $\left\| \cdot \right\|_{F}$ is the Frobenius norm. The solution is given by 
\begin{align}
Q^{(0)}=A_{1:S}G_{0:S-1}^{T}\Big(G_{0:S-1} G_{0:S-1}^{T} \Big)^{-1}.
\label{solutionInitial}
\end{align}
From \eqref{solutionInitial}, we see the justification of the condition $S\ge r$. Namely,  the number of data samples $S$ should satisfy the following condition: $S\ge r$ ($r$ is the number of DM actuators), to ensure that the matrix $G_{0:S-1} G_{0:S-1}^{T}$ in the solution \eqref{solutionInitial} is invertible (the columns of $G_{0:S-1}$ are linearly independent since the control inputs are randomly selected), and consequently, that the solution is well-defined. Once the initial value of the influence function is determined, we determine the initial control actions using the following strategy.

 Namely, the goal of the control algorithm is to produce the desired mirror surface shape, that can be expressed as the vector $\mathbf{a}_{D}\in \mathbb{R}^{M^2}$, consisting of the Walsh coefficients corresponding to the desired mirror shape $W_{D}\in \mathbb{R}^{n\times n}$. Consequently, from \eqref{initialModel}, we can find the set of initial control actions by solving the following optimization problem:
\begin{align}
\min_{\mathbf{g}_{0}} \left\| \mathbf{a}_{D}-Q^{(0)}\mathbf{g}_{0} \right\|_{2}^{2}, \label{optimizationProblemDMInitial1} \\
\text{subject to:}\;\; \mathbf{0} \le \mathbf{g}_{0} \le \mathbf{1}, \label{optimizationProblemDMInitial2}
\end{align}
where the less-than-equal relation operator $\le$ is applied element-wise. If we would simply minimize the cost function \eqref{optimizationProblemDMInitial1} with respect to $\mathbf{g}_{0}$, then most likely the computed control actions would violate the lower limit $0$ and upper limit $1$ on the control actions. Consequently, the constraint \eqref{optimizationProblemDMInitial2} is introduced in order to ensure that the computed control actions are physically realizable. We solve the problem \eqref{optimizationProblemDMInitial1}-\eqref{optimizationProblemDMInitial2} using the interior point method implemented in the MATLAB function lsqlin(). Once the solution of the problem \eqref{optimizationProblemDMInitial1}-\eqref{optimizationProblemDMInitial2} is determined, we can obtain the initial control input vector $u_{i,0}$ for the $i$th actuator as follows 
\begin{align}
u_{i,0}=g_{i,0}^{1/\beta}, \;\;\; i=1,2,\ldots, r,
\label{solution11}
\end{align}
where $g_{i,0}$ is the $i$th entry of $\mathbf{g}_{0}$. The estimated influence matrix $Q^{(0)}$ and the vector $\mathbf{g}_{0}$ are used to initialize the adaptive control method developed in the sequel.

Here, it is important to emphasize that the above-presented approach for generating the initial control actions and the initial influence matrix is actually used for DM control without further improvements, see for example~\cite{Haber:13,Polo:12} and similar least-squares approaches. Our experimental results in Section~\ref{experimentalResultsSection}, show that the developed dual-update control method is able to significantly improve the DM control performance once it is initialized with the generated initial guesses. Consequently, we demonstrate that the developed approach has significant advantages over the least-squares control approaches that are based on a similar strategy to the above-presented strategy used to generate initial guesses.

\subsection{Control Algorithm Development}
In the interest of deriving the control algorithm, we have to introduce one simplification related to the unknown vector $\mathbf{d}_{k}$. In the development of the control algorithm, we assume that the vector $\mathbf{d}_{k}$ does not depend on the control index $k$, that is 
\begin{align}
\mathbf{d}_{k}=\mathbf{d}=\text{const},
\label{assumption1}
\end{align}
Of course, in experiments, this condition might not hold exactly due to the measurement noise. However, this assumption is necessary in order to keep the derivations as simple as possible. As our experiments show, the developed algorithm works very well despite the fact that in experiments there might be some deviations from the condition \eqref{assumption1}. Even without this assumption, it is possible to derive the control algorithm, however, the mathematical apparatus will become more complex and it will involve expectation operators. Taking into account the general model \eqref{mirrorModel}, this assumption gives the following model 
\begin{align}
\mathbf{a}_{k+1} = Q_{k} \mathbf{g}_{k}+\mathbf{d}.
\label{mirrorModel11}
\end{align}
Consequently, to develop the control algorithm, we introduce the control error $\boldsymbol{\varepsilon}_{k} \in \mathbb{R}^{M^2}$ as follows 
\begin{align}
\boldsymbol{\varepsilon}_{k}\coloneqq \mathbf{a}_{D}- \mathbf{a}_{k}.
\label{controlErrorDefinition}
\end{align}
By shifting the time index in  \eqref{controlErrorDefinition} and by combining the resulting equation with  \eqref{mirrorModel11}, we obtain
\begin{align}
\boldsymbol{\varepsilon}_{k+1} = \mathbf{a}_{D}-  Q_{k} \mathbf{g}_{k} - \mathbf{d}.
\label{timeStep1}
\end{align}
On the other hand, for the time step $k$, we have 
\begin{align}
\boldsymbol{\varepsilon}_{k} = \mathbf{a}_{D}-  Q_{k-1} \mathbf{g}_{k-1} - \mathbf{d}.
\label{controlErrorTransform1}
\end{align}
From \eqref{timeStep1} and \eqref{controlErrorTransform1}, we have 
\begin{align}
\boldsymbol{\varepsilon}_{k+1} - \boldsymbol{\varepsilon}_{k} &  =  Q_{k-1} \mathbf{g}_{k-1} -  Q_{k} \mathbf{g}_{k},  \label{controlErrorTransform2} \\
\boldsymbol{\varepsilon}_{k+1} & =  \boldsymbol{\varepsilon}_{k}+ Q_{k-1} \mathbf{g}_{k-1} -  Q_{k} \mathbf{g}_{k},  \label{controlErrorTransform3}\\
\boldsymbol{\varepsilon}_{k+1}& =\boldsymbol{\gamma}_{k}-Q_{k} \mathbf{g}_{k}, \label{controlErrorTransform3}\\
 \boldsymbol{\gamma}_{k}& =   \boldsymbol{\varepsilon}_{k}+ Q_{k-1} \mathbf{g}_{k-1}. \label{gammaDefinition}
\end{align}
Let us assume that at the discrete-time instant $k$, the values of $\boldsymbol{\varepsilon}_{k}$, $Q_{k-1}$, $Q_{k}$, and $\mathbf{g}_{k-1}$ are known. Our goal is to compute the control actions for the next time step $k+1$. That is, our goal is to compute the vector $ \mathbf{g}_{k}$ that consists of the control inputs $u_{i,k}$ (see the equation \eqref{Gvector}). First, we introduce the cost function 
\begin{align}
\boldsymbol{\varepsilon}_{k+1}^{T} \boldsymbol{\varepsilon}_{k+1}=  \Big(  \boldsymbol{\gamma}_{k} -  Q_{k} \mathbf{g}_{k}\Big)^{T}\Big( \boldsymbol{\gamma}_{k}-  Q_{k} \mathbf{g}_{k}\Big).
\label{costFunction}
\end{align}
Similarly to the optimization problem in \eqref{optimizationProblemDMInitial1}-\eqref{optimizationProblemDMInitial2}, we compute the control inputs for the time instant $k+1$, by solving the following constrained optimization problem 
\begin{align}
& \min_{\mathbf{g}_{k}}  \Big(  \boldsymbol{\gamma}_{k} -  Q_{k} \mathbf{g}_{k}\Big)^{T}\Big( \boldsymbol{\gamma}_{k}-  Q_{k} \mathbf{g}_{k}\Big), \label{optimization1}  \\
& \text{subject to}\;\; \mathbf{0} \le  \mathbf{g}_{k} \le \mathbf{1}.  \label{constraint1}
\end{align}
We solve the problem \eqref{optimization1}-\eqref{constraint1} using the interior point method implemented in the MATLAB function lsqlin(). Once the solution, denoted by $\mathbf{g}_{k}$, of the problem \eqref{optimization1}-\eqref{constraint1} is determined, we can obtain the control input vector $u_{i,k}$ for the $i$th actuator as follows 
\begin{align}
u_{i,k}=g_{i,k}^{1/\beta},
\label{solution112}
\end{align}
where $g_{i,k}$ is the $i$th entry of $\mathbf{g}_{k}$.

\subsection{Dual-update control algorithm summary}

To formulate and solve the problem \eqref{optimization1}-\eqref{constraint1}, we need to compute the quantities $\boldsymbol{\varepsilon}_{k}$, $Q_{k-1}$, and  $\mathbf{g}_{k-1}$ that form the vector $\boldsymbol{\gamma}_{k}$, as well as the matrix $Q_{k}$. We use a recursive identification approach to compute these quantities~\cite{ljung1999}. To define the recursive identification approach, we need to define the following quantities. By applying the $\text{vec}(\cdot)$ operator to \eqref{mirrorModel11}, we obtain 
  \begin{align}
\mathbf{a}_{k+1}& =G_{k} \mathbf{q}_{k} +\mathbf{d}, \label{mirrorModel11Vec1} \\
G_{k}& =\mathbf{g}_{k}^{T}\otimes I, \label{mirrorModel11Vec2} \\
\mathbf{q}_{k}& = \text{vec}\big(Q_{k} \big) \label{mirrorModel11Vec3},
 \end{align}
  where $\mathbf{q}_{k} \in \mathbb{R}^{rM^{2}}$ is the vector that parametrizes the influence matrix, $G_{k}\in \mathbb{R}^{M^2 \times rM^2}$, and we have used the following property of the $\text{vec}(\cdot)$ operator~\cite{haber2014sparseLyapunov}: $\text{vec}\big(X_{1}X_{2}X_{3}  \big)=\big(X_{3}^{T}\otimes X_{1} \big)\text{vec}(X_{2})$. 

The dual-update control method consists of the following steps.
\\\\
\textbf{Step 1:} At time step $k$, the following values are available from the previous steps: $\mathbf{a}_{k}$, $Q_{k-1}$, and $\mathbf{g}_{k-1}$. Using these values, compute  $\boldsymbol{\varepsilon}_{k}$ given by \eqref{controlErrorDefinition} and $\boldsymbol{\gamma}_{k}$ given by \eqref{gammaDefinition}. Form and solve the optimization problem \eqref{optimization1}-\eqref{constraint1}, using the MATLAB function lsqlin(). The solution is given by $\mathbf{g}_{k}$. Using this value, compute the control actions \eqref{solution11}. Apply the control actions to the DM.
\\ \\
\textbf{Step 2:} Wait for the time instant $k+1$, and observe the mirror surface response $\mathbf{w}_{k+1}$. Use the mirror surface observation to compute the vector $\mathbf{a}_{k+1}$ (observed coefficients) using \eqref{projection22} (for the shifted time index $k+1$). On the basis of the computed value $\mathbf{g}_{k}$, from the previous step $k$, form the matrix $G_{k}$ using \eqref{mirrorModel11Vec2}.  
 The following values are available from the previous time step $k$: $P_{k} \in \mathbb{R}^{rM^2 \times rM^2}$ and $\mathbf{q}_{k}$. Update the matrix $P_{k}$ and the vector of parameters $\mathbf{q}_{k}$ by performing the following steps:
  \begin{align}
S_{k}&=\Big(\lambda I +G_{k}P_{k} G_{k}^{T} \Big)^{-1}, \label{recursiveIdentification1} \\
L_{k+1}&=P_{k}G_{k}^{T}S_{k}, \label{recursiveIdentification2} \\
P_{k+1}&=\frac{1}{\lambda} P_{k}-\frac{1}{\lambda} L_{k+1} G_{k} P_{k}, \label{recursiveIdentification3} \\
\mathbf{e}_{k+1}&=\mathbf{a}_{k+1}-G_{k}\mathbf{q}_{k}, \label{recursiveIdentification31}\\
\mathbf{q}_{k+1}&=\mathbf{q}_{k}+L_{k+1}\mathbf{e}_{k+1}. \label{recursiveIdentification41}
\end{align}
where $0<\lambda \le 1$ is a user selected parameter, $\mathbf{e}_{k+1}\in \mathbb{R}^{M^2}$, $S_{k}\in \mathbb{R}^{M^2 \times M^2}$, and $L_{k+1}\in \mathbb{R}^{rM^2\times M^2}$ is the gain matrix. If the convergence of the control error is not achieved, go to Step 1 (index in Step 1 now becomes k+1), otherwise, stop the recursion. 
\\
\\
Several comments about the developed algorithm are in order. The vector $\mathbf{e}_{k+1}$ defined in \eqref{recursiveIdentification31} is called the model error. This vector quantifies the difference between the observed modal response $\mathbf{a}_{k+1}$ and the model prediction $G_{k}\mathbf{q}_{k}$. Following the guidelines given in~\cite[p.~379]{ljung1999} and \cite[p.~68]{landau2011adaptive}, we use $\lambda = 0.98$. However, other possibilities for selecting $\lambda$ are also possible~\cite{landau2011adaptive,ljung1983theory}. The matrix $P_{k}$ is initialized as $P_{0}=0.05 I$. Here we have used a scaled identity matrix (sparse matrix) to initialize $P_{k}$, in order to minimize the computational burden. This is necessary since the matrices in \eqref{recursiveIdentification1}-\eqref{recursiveIdentification41} are large dimensional. Namely, initialization of $P_{k}$ as a dense matrix will significantly increase the computational burden. The issue of decreasing the computational complexity of the steps \eqref{recursiveIdentification1}-\eqref{recursiveIdentification41} is a future research topic. The quantities $\mathbf{g}_{k}$ and $Q_{k}$ are initialized with the values $\mathbf{g}_{0}$ and $Q^{(0)}$ computed in the initialization step explained in Section~\ref{initialEstimation}.

\section{Experimental Results}
\label{experimentalResultsSection}

In this section, we present the experimental results.

We test the developed approach using a Boston Micromachines MEMS DM with $r=140$ actuators. The actuation grid is $12$ by $12$ with all $4$ corner actuators being inactive. The DM stroke is $3.5$ $[\mu m ]$ and the pitch is $400$ $[\mu m]$. The behavior of this  DM type has been analyzed in many papers, see for example~\cite{diouf2010open,stewart2007open} and follow-up works. Consequently, due to paper brevity, we do not further summarize other mirror properties.

The produced mirror surface shape is sensed by a Partitioned Aperture Wave-front (PAW) sensor ~\cite{barankov2013single,parthasarathy2012quantitative,li2015conjugate}. This sensor has a large dynamic range, it is relatively fast, and it operates with uncollimated light sources. It has a relatively high resolution that is only limited by the used camera. In addition, this sensor is speckle-free, robust, and polarization-independent. Further details related to this sensor can be found in~\cite{parthasarathy2012quantitative,li2015conjugate}. The used experimental setup is the same as the experimental setup used to generate the results in our previous publication~\cite{haber2021modeling}. Consequently, in the interest of brevity, we just mention its basic configuration. The light source is a red LED (660 [nm], Thorlabs). A system of optical components is used to direct and shape the beam and to illuminate the DM surface. A monochrome camera (Blackfly BFS-U3-123S6M-C) is used as a PAW sensing element. The maximal size of the observed image (deformation matrix) is $1001$ by $999$. The DM and sensor are controlled using the MATLAB programming environment. 

As explained in Section~\ref{WalshSection}, to decompose the observed deformation as a sum of Walsh pattern matrices, the deformation matrix size should be expressed as a power of $2$. On the other hand, the camera of the PAW sensor produces a deformation image size of $1001$ by $999$. This image covers an area that is larger than the active DM area. Taking all these things into consideration, we have at least two options for selecting the deformation matrix size. The first option is $256$ by $256$ matrix, and the second option is $512$ by $512$. The second option covers the centers of all the actuators (including $4$ corner inactive actuators). However, in the second option, a part of deformation caused by the edge actuators will not take part in the defined $512$ by $512$ deformation matrix. We investigate the performance of the developed method for both options. 

\subsection{Results for the 256 by 256 deformation matrix}

\begin{figure}[H]
\centering 
\includegraphics[scale=0.48,trim=0mm 0mm 0mm 0mm ,clip=true]{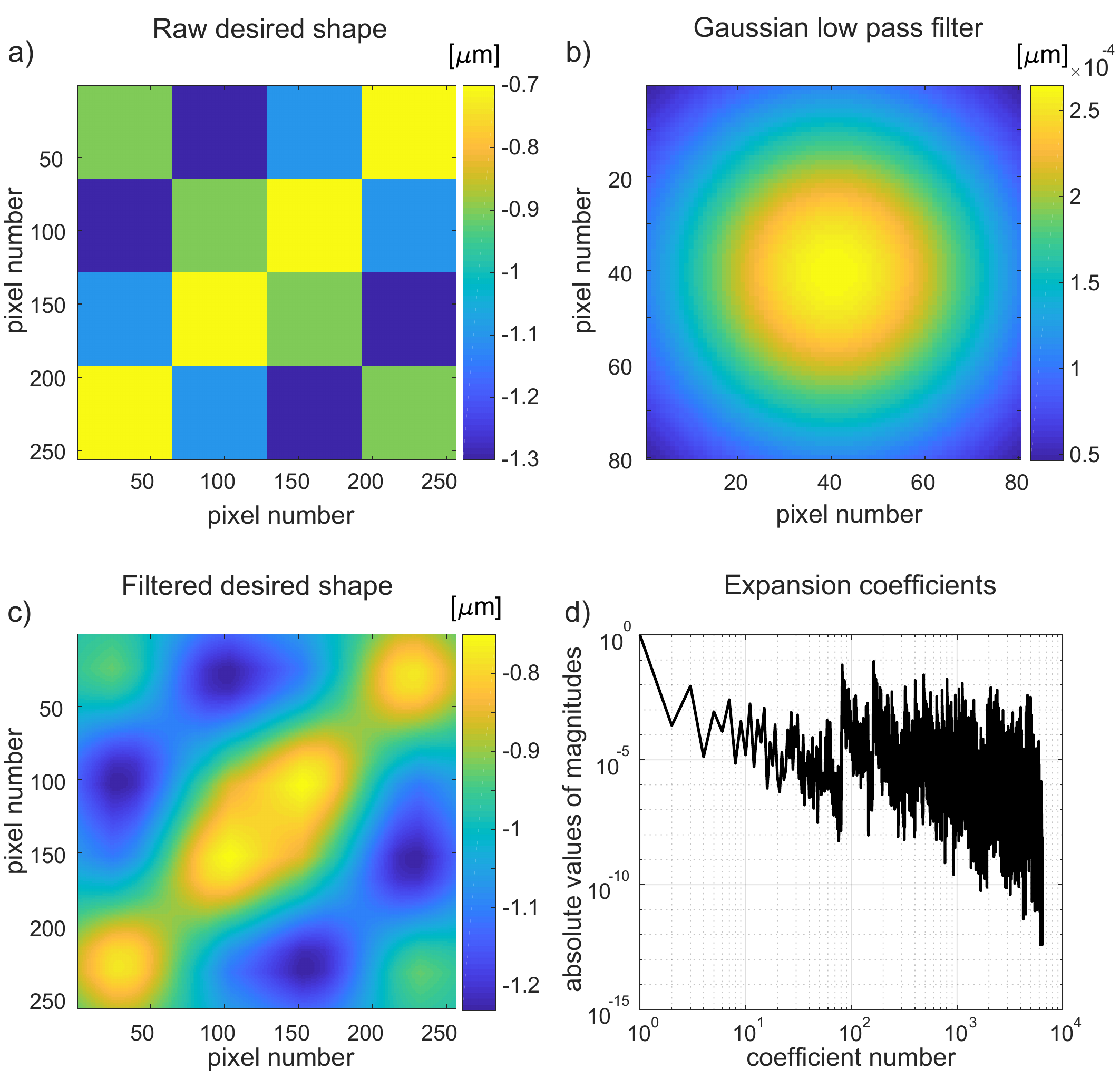}
\caption{(a) ``Raw'' desired shape defined in \eqref{rawSurfaceShape1}. (b) Gaussian 2D low-pass filter applied to the ``raw'' desired shape. (c) Desired shape after applying the filter and offset. (d) Coefficients of the decomposition \eqref{matrixDecomposition} obtained by decomposing the filtered desired shape.}
\label{fig:Graph2}
\end{figure}

First, we present the control results for the active-controlled surface represented by $256$ by $256$ image ($256$ by $256$ deformation matrix). This corresponds to $n=256=2^{8}$, that is, to $V=8$ (for more details, see Section~\ref{WalshSection}). We generate a ``raw'' desired mirror surface shape as
\begin{align}
W_{D}^{\text{raw}}= a_{1,1}Z_{1,1}+a_{2,2}Z_{2,2}+a_{3,3}Z_{3,3},
\label{rawSurfaceShape1}
\end{align}
with the coefficients of the expansion equal to $a_{1,1}=-1$, $c_{2,2}=-0.1$, and $c_{3,3}=0.2$. The raw desired surface is shown in Fig.~\ref{fig:Graph2}(a). This desired surface contains vertical (90 degree) surface changes between the segments of the regular checkerboard pattern shown in Fig.~\ref{fig:Graph2}(a). 
On the other hand, the deformation response of a single actuator is a smooth function resembling the Gaussian function. Consequently, the mirror actuators are not able to produce vertical 90 degree deformation changes from one segment to another. Consequently, we need to apply a smoothing low-pass 2D filter to the desired raw surface shape. We choose the Gaussian 2D filter shown in Fig.~\ref{fig:Graph2}(b). The standard deviation of the filter is $30$. After applying the Gaussian 2D filter, we offset the resulting deformation by $-1$ (from every entry of the matrix we subtract $-1$). This process produces the desired surface shape shown in Fig.~\ref{fig:Graph2}(c). We compute the decomposition given by the equation~\eqref{matrixDecomposition} for $M=80$. This produces a total of $6400$ coefficients that are shown in Fig.~\ref{fig:Graph2}(d). These are the desired coefficients that we want to produce.

\begin{figure}[H]
\centering 
\includegraphics[scale=0.48,trim=0mm 0mm 0mm 0mm ,clip=true]{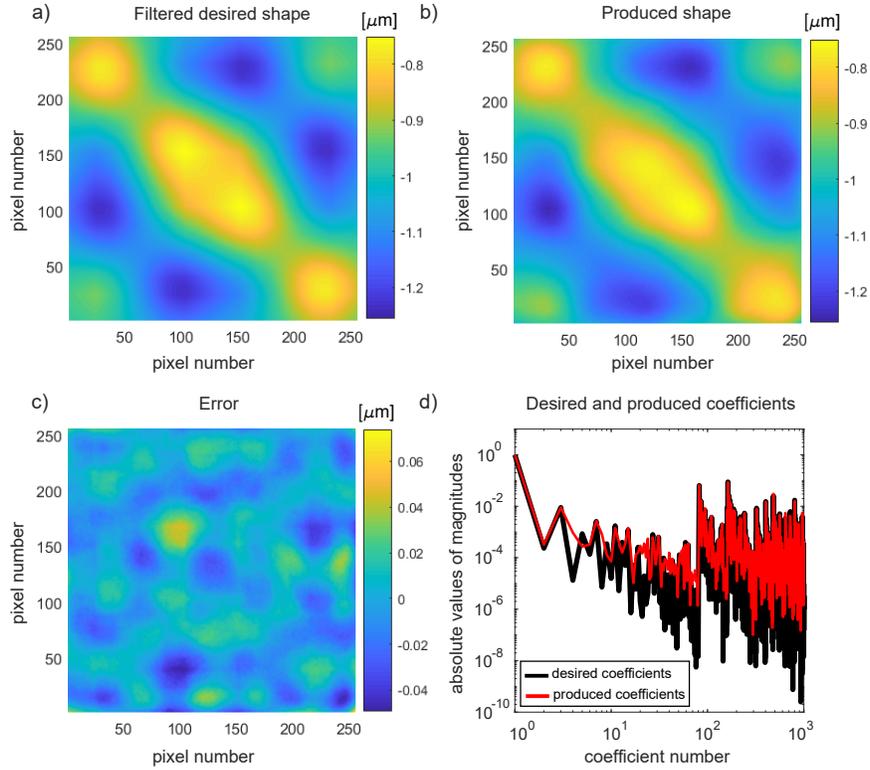}
\caption{(a) Desired filtered mirror shape. (b) Best produced mirror shape. (c) Error. (d) Desired and produced coefficients.}
\label{fig:Graph3}
\end{figure}

Next, we compute the control actions using the developed approach. Figure~\ref{fig:Graph3} shows the control results. Panels (a) and (b) in Fig.~\ref{fig:Graph3} show the desired and best-produced shapes, respectively. Panel (c) in Fig.~\ref{fig:Graph3} shows the error (difference between the desired and produced shapes). The Root-Mean-Square (RMS) surface error is $14$ $[nm]$. Finally, Fig.~\ref{fig:Graph3}(d) shows the desired and best-produced coefficients of the 2D Walsh basis expansion~\eqref{matrixDecomposition}. 

\begin{figure}[H]
\centering 
\includegraphics[scale=0.48,trim=0mm 0mm 0mm 0mm ,clip=true]{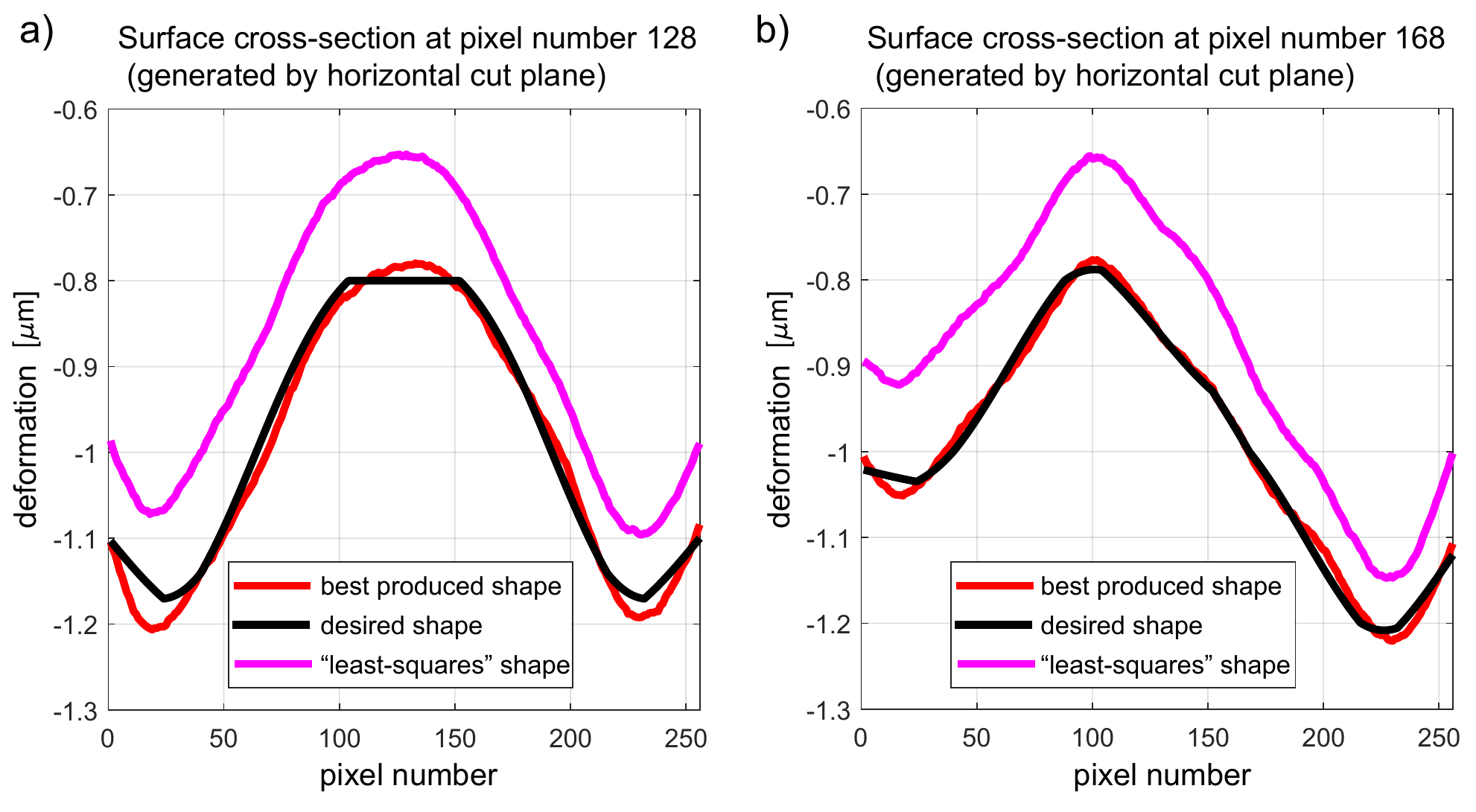}
\caption{Surface cross sections at the horizontal cut planes generated at the pixels  (a) $128$  and  (b) $168$.  The ``least-squares'' shape corresponds to the initial surface shape produced by the initial control actions generated by solving \eqref{optimizationProblemDMInitial1}-\eqref{optimizationProblemDMInitial2}.}
\label{fig:Graph4}
\end{figure}

\begin{figure}[H]
\centering 
\includegraphics[scale=0.48,trim=0mm 0mm 0mm 0mm ,clip=true]{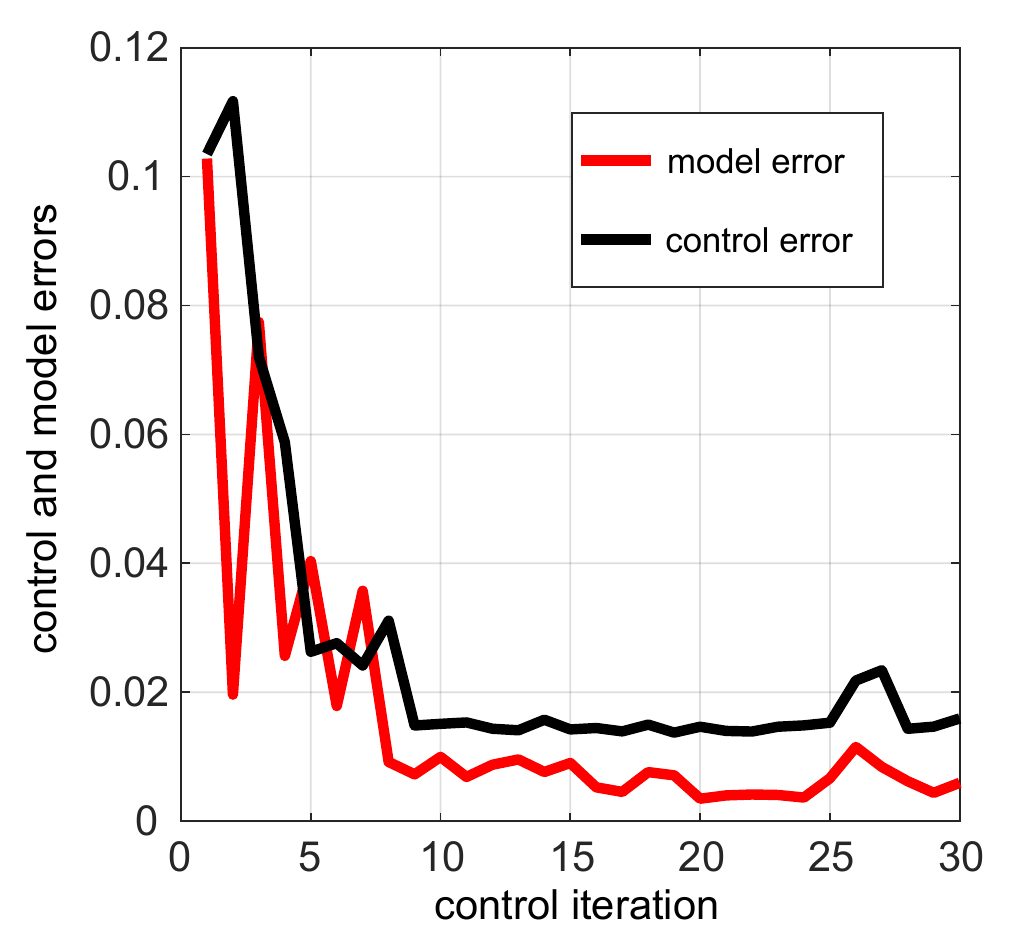}
\caption{Convergence of the control method. The graphs shows the 2-norms of the control error $\boldsymbol{\varepsilon}_{k+1}$  and model error  $\mathbf{e}_{k+1}$ defined in \eqref{controlErrorDefinition} and \eqref{recursiveIdentification31}, respectively.}
\label{fig:Graph5}
\end{figure}

Figure~\ref{fig:Graph4} shows the surface cross section generated at two horizontal cutting planes. Figure~\ref{fig:Graph5} shows the convergence of the control error $\boldsymbol{\varepsilon}_{k+1}$ defined in \eqref{controlErrorDefinition} and the  model error $\mathbf{e}_{k+1}$ defined in \eqref{recursiveIdentification31}.

The presented results demonstrate the excellent performance of the developed method. The RMS surface error converges to $14.1$ $[nm]$ in a relatively small number of iterations. Furthermore, from Fig.~\ref{fig:Graph4} we can observe that the method outperforms the ``least-squares'' approach for controlling the DM. In our simulations, this approach, which is explained in Section~\ref{initialEstimation}, is used to generate the initial guess. However, this approach is used in a number of articles to control the DM. That is, the adaptive control method, proposed in this article, clearly has a significant advantage over the classical least-squares approach for DM control.

\subsection{Results for the 512 by 512 deformation matrix}

\begin{figure}[H]
\centering 
\includegraphics[scale=0.48,trim=0mm 0mm 0mm 0mm ,clip=true]{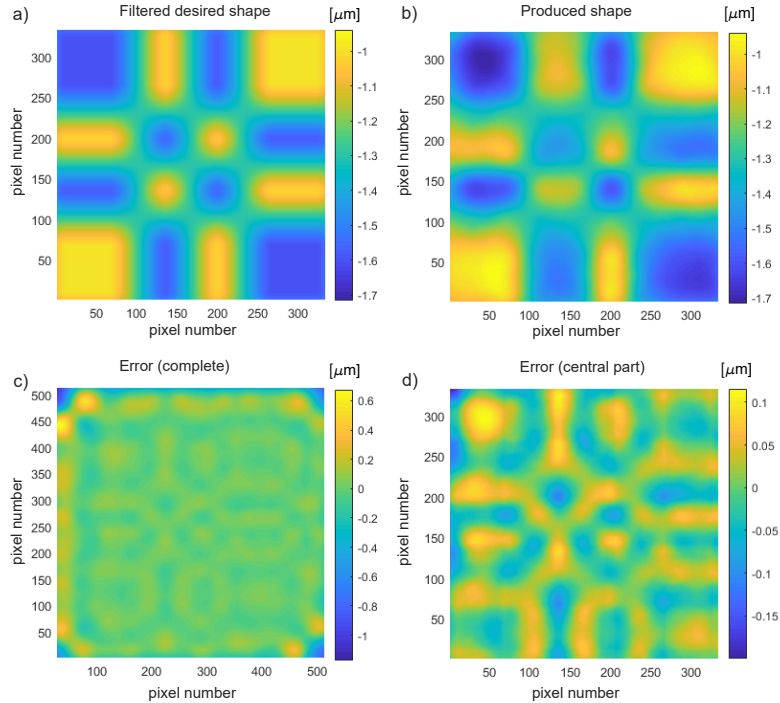}
\caption{Control result for the raw desired shape \eqref{desiredShape1} (note that this shape is filtered and scaled to generate panel (a)). (a) Filtered and scaled desired shape. (b) Best produced shape. (c) Surface error. (d) Surface error for the central mirror part (obtained by cropping the surface error shown in panel (c) by 90 pixels on all 4 image sides).}
\label{fig:Graph6}
\end{figure}

\begin{figure}[H]
\centering 
\includegraphics[scale=0.48,trim=0mm 0mm 0mm 0mm ,clip=true]{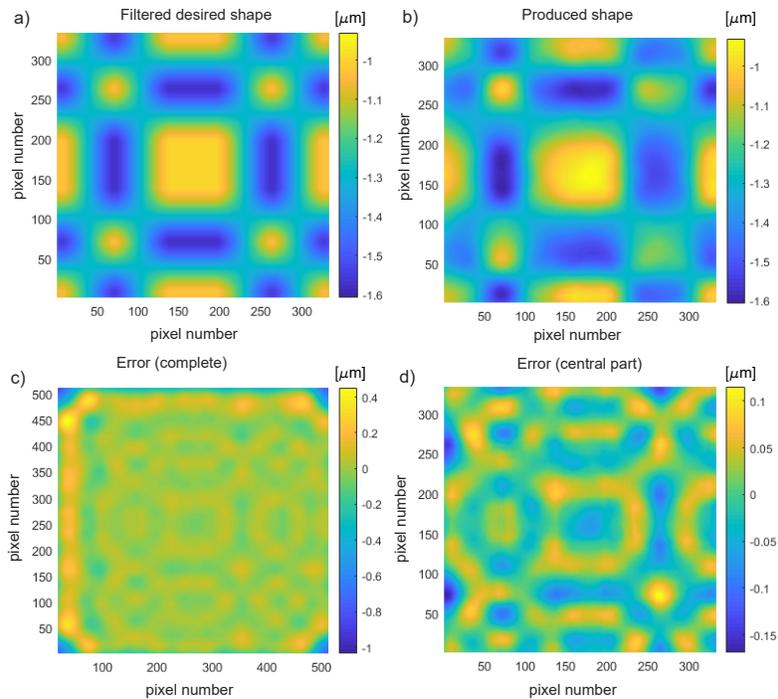}
\caption{Control result for the raw desired shape \eqref{desiredShape2} (note that this shape is filtered and scaled to generate panel (a)). Captions of panels in this figure correspond to the captions of panels in Fig.~\ref{fig:Graph6}.}
\label{fig:Graph8}
\end{figure}

\begin{figure}[H]
\centering 
\includegraphics[scale=0.48,trim=0mm 0mm 0mm 0mm ,clip=true]{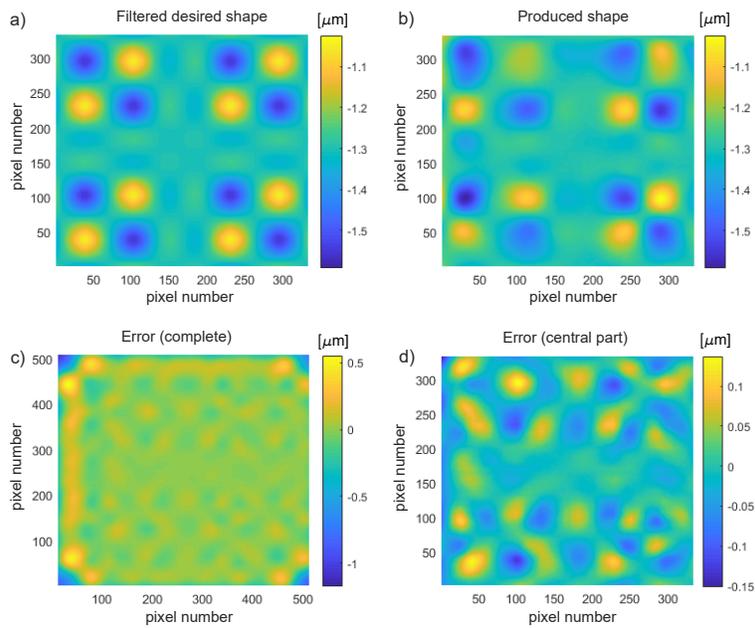}
\caption{Control result for the raw desired shape \eqref{desiredShape3} (note that this shape is filtered and scaled to generate panel (a)). Captions of panels in this figure correspond to the captions of panels in Fig.~\ref{fig:Graph6}.}
\label{fig:Graph10}
\end{figure}

Here we present the results for the observed deformation matrix with the dimension of $512$ by $512$ pixels. In this case $n=2^{V}$, where $v=9$, for more details, see Section~\ref{WalshSection}. By using this choice of the deformation matrix, we are able to investigate the influence of the actuation boundaries on the performance of the developed algorithm. We test the following raw desires shapes with spatial frequencies from smaller to larger:
\begin{align}
W_{D1}=-0.3 Z_{1,1}+ 0.3 Z_{6,6}, \label{desiredShape1}\\
W_{D2}=-0.3 Z_{1,1}+ 0.3 Z_{7,7}, \label{desiredShape2} \\
W_{D3}=-0.3 Z_{1,1}+ 0.3 Z_{10,10}. \label{desiredShape3}
\end{align}
We apply the Gaussian low-pass filter to the desired shapes. The filter has the support of $70$ pixels and deviation of $25$. Once the filter is applied, on offset of $-1$ is applied to the filtered shapes to produce the final desired shapes. The final desired shapes are shown in panels (a) of Figs.~\ref{fig:Graph6},~\ref{fig:Graph8}, and~\ref{fig:Graph10}. Panels (b) in Figs.~\ref{fig:Graph6},~\ref{fig:Graph8}, and~\ref{fig:Graph10}, show the best-produced shapes. Panels (c) in Figs.~\ref{fig:Graph6},~\ref{fig:Graph8}, and~\ref{fig:Graph10}, show the surface error. Panels (d) in Figs.~\ref{fig:Graph6},~\ref{fig:Graph8}, and~\ref{fig:Graph10}, show the surface error over the central mirror part (obtained by cropping the surface error by 90 pixels on all 4 image sides). 

\begin{figure}[H]
\centering 
\includegraphics[scale=0.48,trim=0mm 0mm 0mm 0mm ,clip=true]{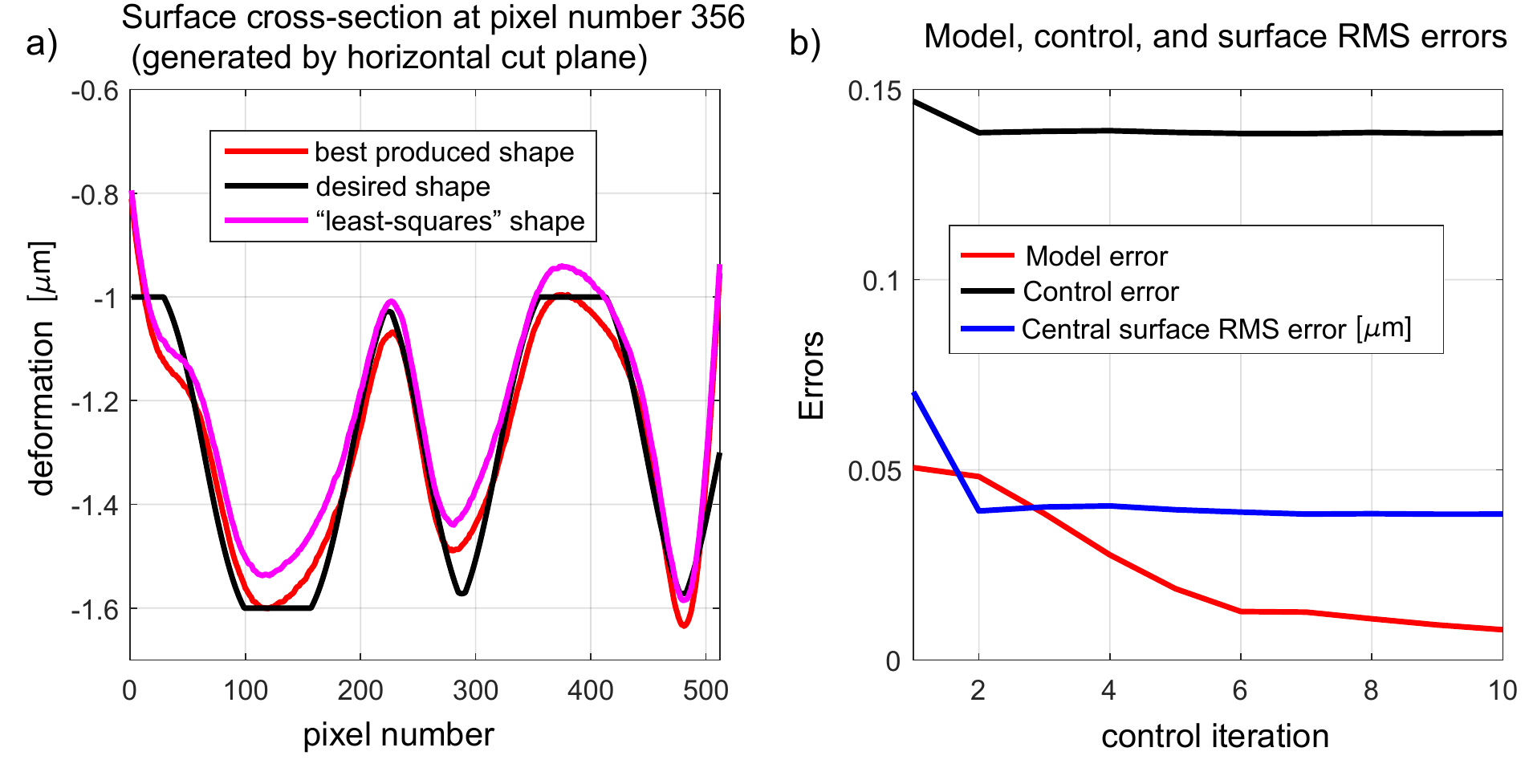}
\caption{Control result for the raw desired shape \eqref{desiredShape1} (note that this shape is filtered and scaled to generate panel (a) in Fig.~\ref{fig:Graph6}). (a) Cross-sections of the produced and desired shapes generated by horizontal cut planes. The ``least-squares'' shape corresponds to the initial surface shape produced by the initial control actions generated by solving \eqref{optimizationProblemDMInitial1}-\eqref{optimizationProblemDMInitial2}. (b) Convergence of the model, control, and RMS surface errors of the developed algorithm. Panel (b)  shows the 2-norms of the control error $\boldsymbol{\varepsilon}_{k+1}$  and model error  $\mathbf{e}_{k+1}$ defined in \eqref{controlErrorDefinition} and \eqref{recursiveIdentification31}, respectively.}
\label{fig:Graph7}
\end{figure}

\begin{figure}[H]
\centering 
\includegraphics[scale=0.48,trim=0mm 0mm 0mm 0mm ,clip=true]{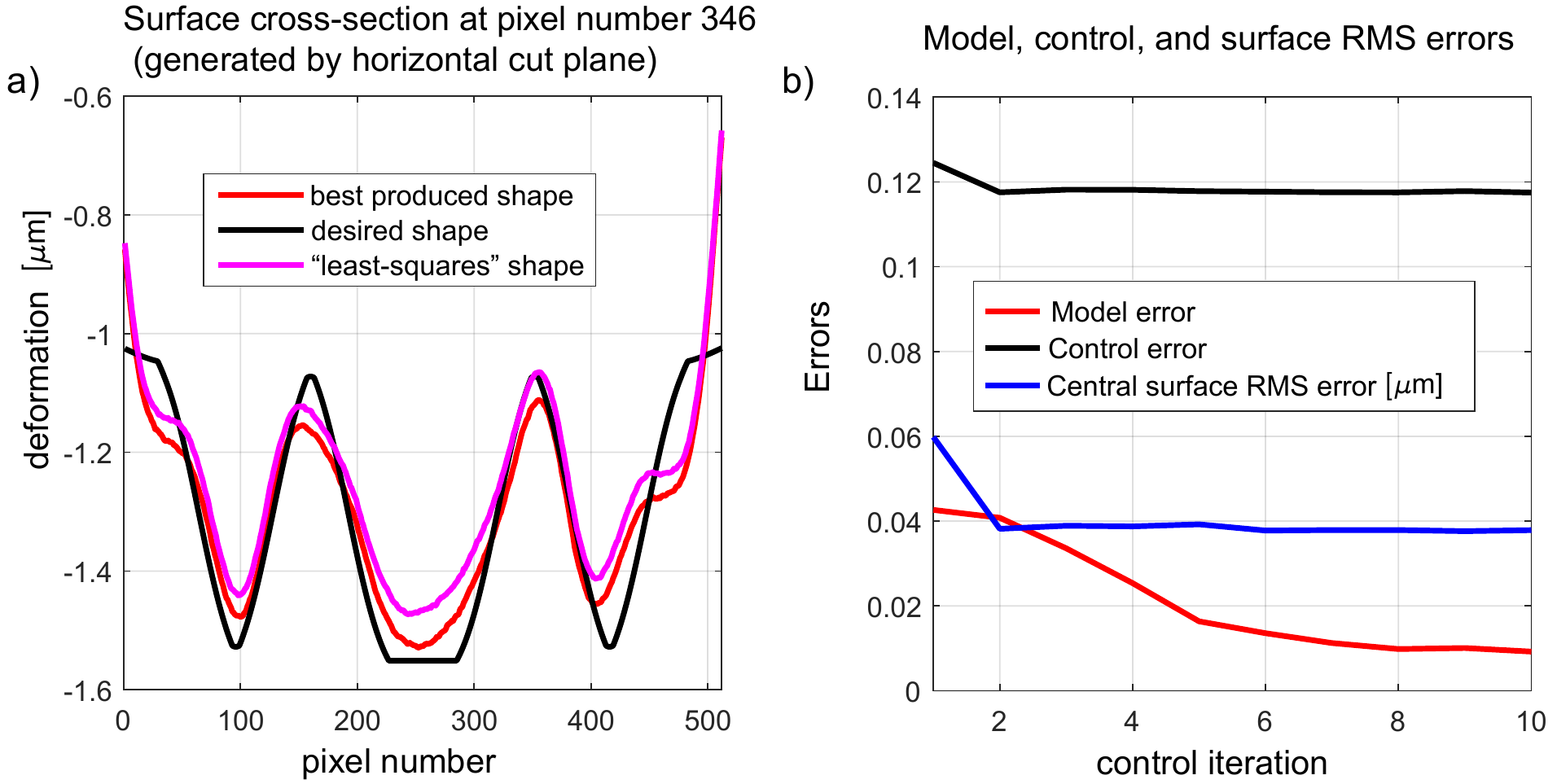}
\caption{Control result for the raw desired shape \eqref{desiredShape2} (note that this shape is filtered and scaled to generate panel (a) in Fig.~\ref{fig:Graph8}). Captions of panels in this figure correspond to the captions of panels in Fig.~\ref{fig:Graph7}.}
\label{fig:Graph9}
\end{figure}

\begin{figure}[H]
\centering 
\includegraphics[scale=0.48,trim=0mm 0mm 0mm 0mm ,clip=true]{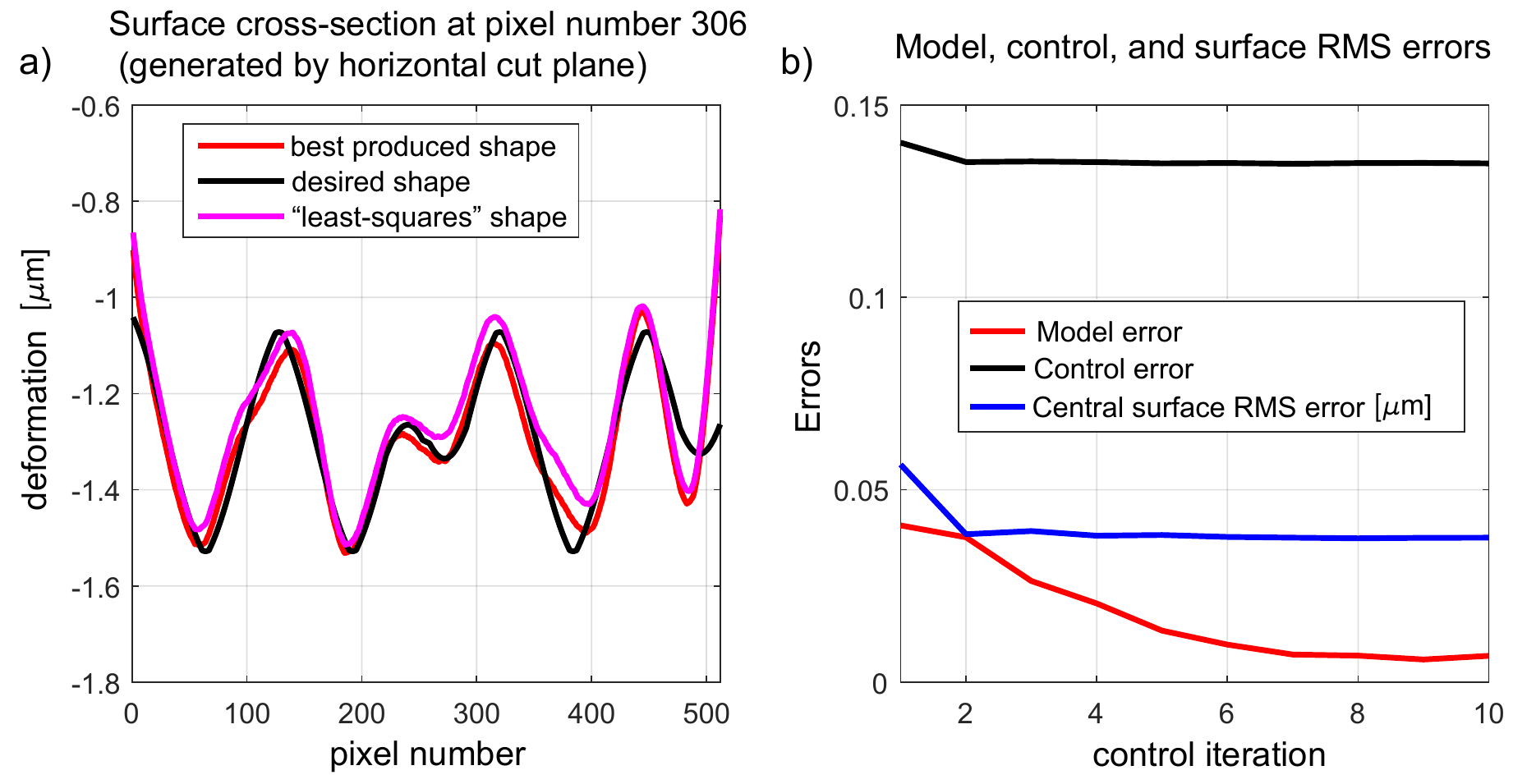}
\caption{Control result for the raw desired shape \eqref{desiredShape3} (note that this shape is filtered and scaled to generate panel (a) in Fig.~\ref{fig:Graph10}). Captions of panels in this figure correspond to the captions of panels in Fig.~\ref{fig:Graph7}.}
\label{fig:Graph11}
\end{figure}

Panels (a) in Figs.~\ref{fig:Graph7},~\ref{fig:Graph9}, and~\ref{fig:Graph11}, show the cross-sections of the desired and produced shapes generated by horizontal cut planes.  Panels (b) in Figs.~\ref{fig:Graph7}, ~\ref{fig:Graph9}, and~\ref{fig:Graph11} show the convergence of the control and model errors. The above-presented results demonstrate the excellent performance of the developed method.  From panels  (b) in Figs.~\ref{fig:Graph7}, ~\ref{fig:Graph9}, and~\ref{fig:Graph11}, we can observe that the RMS surface error converges to approximately $40$ $[nm]$ in a relatively small number of iterations. These results can additionally be improved by tuning the parameters (parameter $\lambda$ and the matrix $P$) of the algorithm. The development of methods for optimal tuning of the dual-update algorithm is a future research topic. Furthermore, by comparing the ``least-squares'' shapes with the desired shapes in panels (a) of the same figures, we can observe that our method has significant advantages over the state-of-the-art least-squares approaches for DM control.

\section{Conclusion}
\label{SectionConclusion}

In this paper, we have developed a novel approach for adaptive control of DMs. On the basis of the feedback information provided by the sensor, our method updates both the DM influence matrix and the control actions. In this way, we are able to generate RMS control accuracy of $14-40$ $[nm]$.  These results can additionally be improved by tuning the parameters of the algorithm.  Besides introducing a novel control method, we have also demonstrated a good potential of using Walsh basis functions for DM control. Walsh basis function can achieve its full potential for control of segmented DMs. Our approach can straightforwardly be applied to the control problems involving segmented DMs. In future work, we will focus on improving the performance of the developed approach by optimally tuning the control algorithm parameters, and on reducing the computational complexity of the developed approach. 

\bibliographystyle{unsrt}
\bibliography{sample}
\end{document}